\documentclass[final,5p,times,twocolumn]{elsarticle}
\biboptions{numbers,sort&compress}

\usepackage{amssymb}
\usepackage{todonotes,soul}
\usepackage{geometry}
\usepackage{mathrsfs}
\usepackage{amsthm,amsmath}
\usepackage{tikz}
\usepackage{float}
\usepackage{enumitem}
\usepackage{physics}
\usepackage{dsfont}
\usepackage{mathtools}
\usepackage{subcaption}
\usepackage{url}
\usepackage{amsmath, bm}
\usepackage{gensymb}
\usepackage{amssymb}
\usepackage{xfrac}
\usepackage{dirtytalk}
\usepackage{pgfplots}
\pgfplotsset{width = 10cm, compat = 1.9}
\usepackage{cancel}
\usepackage{tensor}
\usepackage{siunitx}
\usepackage[nottoc]{tocbibind}
\usepackage{xcolor}
\usepackage{hyperref}
\usepackage{sidecap}
\numberwithin{equation}{section}

\allowdisplaybreaks

\begin{document}

\begin{frontmatter}

\title{Quantum effects in rotating thermal states on anti-de Sitter space-time}

\author[1]{Jacob C. Thompson} 
\ead{JThompson16@sheffield.ac.uk}
\author[1]{Elizabeth Winstanley\corref{cor1}}
\ead{E.Winstanley@sheffield.ac.uk}

\affiliation[1]{organization={School  of Mathematical and Physical Sciences, The University of Sheffield},
            addressline={Hicks Building, Hounsfield Road}, 
            city={Sheffield},
            postcode={S3 7RH}, 
            country={United Kingdom}}

\cortext[cor1]{Corresponding author}

\begin{abstract}
We study the stress-energy tensor of a massless, conformally-coupled, quantum scalar field in a rigidly-rotating thermal state on three- and four-dimensional anti-de Sitter space-time.
We first find the stress-energy tensor using relativistic kinetic theory, modelling the field as a thermal gas of massless bosons. 
We then compute the renormalized stress-energy tensor of the scalar field in quantum field theory and compare it with that resulting from relativistic kinetic theory.
\end{abstract}

\end{frontmatter}

\section{Introduction}
\label{sec:intro}

Anti-de Sitter (AdS) space-time is a maximally-symmetric solution of the Einstein equations having constant negative scalar curvature, of particular importance for holography and the AdS/CFT correspondence \cite{Aharony:1999ti}. 
The simplicity of this space-time geometry facilitates the study of quantum field theory (QFT) on this background, rendering AdS a useful prototype for exploring QFT effects on more general curved space-times,
while nonetheless giving nontrivial results.

A global vacuum state can be defined on AdS \cite{Avis:1977yn}, in analogy with the global Minkowski vacuum state, and both of these preserve the maximal symmetry of the underlying space-time.
In contrast, while a global thermal state on Minkowski space-time also inherits maximal symmetry, this is not the case for a global thermal state on AdS \cite{Allen:1986ty}.
Due to the curvature of AdS, thermal radiation tends to ``clump'', and is localized near  the origin \cite{Allen:1986ty,Ambrus:2018olh}.

Our focus in this letter is rigidly-rotating thermal states.
No rigidly-rotating vacuum or thermal states can be defined for a quantum scalar field on unbounded Minkowski space-time \cite{Duffy:2002ss}.
For a quantum fermion field, rigidly-rotating thermal states can be defined, but the latter are regular 
only inside the speed-of-light surface (the surface outside which rigidly-rotating observers must be travelling at speeds greater than the speed of light) \cite{Ambrus:2014uqa}. 
Regular rigidly-rotating thermal states on Minkowski space-time can be constructed for both bosonic and fermionic quantum fields if the space-time is bounded by a perfectly reflecting mirror inside the speed-of-light surface \cite{Duffy:2002ss,Ambrus:2015lfr,Zhang:2020hct}, but at the cost of introducing Casimir-like divergences on the mirror \cite{Duffy:2002ss,Ambrus:2015lfr}.

The global structure of AdS space-time provides a natural boundary, and thus rigidly-rotating vacuum and thermal states can be defined without recourse to an artificial mirror (thus avoiding the consequent Casimir divergences), providing that the rate of rotation is sufficiently small that there is no speed-of-light surface.
In this case, rigidly-rotating thermal states have been extensively studied for fermions \cite{Ambrus:2014fka,Ambrus:2021eod}, but no similar study for a quantum scalar field has been performed to date. 

In this letter we investigate the properties of the 
renormalized stress-energy tensor (RSET) for a massless, conformally-coupled quantum scalar field in a rigidly-rotating thermal state on AdS, assuming that there is no speed-of-light surface. 
For fermions \cite{Ambrus:2014fka}, when the rate of rotation is sufficiently small, thermal radiation again ``clumps'' in a neighbourhood of the origin, but increasing the rate of rotation results in an increasingly energetic thermal state, with the location of the maximum energy density moving from the origin outwards towards the space-time boundary.
One purpose of our analysis is to explore whether a similar effect occurs for a quantum scalar field.

We also compare our QFT results with those arising from relativistic kinetic theory (RKT), in which the quantum scalar field is modelled as a relativistic gas of massless classical bosonic particles.   
For nonrotating thermal states, RKT yields results which are identical to those in QFT on Minkowski space-time \cite{Ambrus:2017vlf,Ambrus:2018olh}, while on AdS quantum effects give small corrections to the RKT results, the relative magnitude of the corrections decreasing as the temperature increases \cite{Ambrus:2017vlf,Ambrus:2018olh}.
Away from the reflecting mirror, RKT also gives a good approximation to the RSET for a quantum scalar field in a rigidly-rotating thermal state on bounded Minkowski space-time \cite{Duffy:2002ss}.

The outline of this letter is as follows.
In Sec.~\ref{sec:rotAdS} we describe the metric for rigidly-rotating AdS space-time in global coordinates, working in three and four space-time dimensions.
The SET (stress-energy tensor) is derived using RKT in Sec.~\ref{sec:RKT}, followed by the corresponding QFT construction in Sec.~\ref{sec:QFT}.
In Sec.~\ref{sec:QC} we compare our results from RKT and QFT.
Finally, Sec.~\ref{sec:conc} contains our conclusions.

\section{Rotating AdS space-time}
\label{sec:rotAdS}

AdS space-time in global coordinates in $n$ space-time dimensions has line element
\begin{subequations}
\label{eq:metric}
\begin{equation}
    {\mathrm {d}}s^{2} = a^{-2} \sec ^{2}\rho \, \left[  -{\mathrm {d}}\tau ^{2} + {\mathrm {d}}\rho ^{2} + \sin ^{2} \rho \, {\mathrm {d}}S^{2}_{n-2} \right] ,
\end{equation}
where ${\mathrm {d}}S^{2}_{n-2}$ is the metric on the $(n-2)$-dimensional sphere and $a$ is the inverse AdS radius of curvature. 
Throughout this paper, we use units in which $c=G=\hbar = k_{\mathrm {B}}=1$.
The coordinate ranges are $-\infty < \tau < \infty  $ and $0\le \rho < \pi /2$.
For simplicity, we focus in this paper on three- and four-dimensional AdS, for which cases we have
\begin{equation}
    {\mathrm {d}} S^{2}_{n-2} = \begin{cases}
    {\mathrm {d}}{\widetilde {\varphi }}^{2}, & n=3 ,  \\
    {\mathrm {d}}\theta ^{2} + \sin ^{2} \theta \, {\mathrm {d}}{\widetilde {\varphi }}^{2}, &  n=4 ,
    \end{cases}
\end{equation}
\end{subequations}
where $0\le  \theta \le \pi $ and $0 \le {\widetilde {\varphi }}< 2\pi  $.

We are interested in states which are rigidly-rotating with angular speed $\Omega $ in the ${\widetilde {\varphi }}$ direction.
We therefore define a corotating angle $\varphi $ by
    $\varphi  = {\widetilde {\varphi }} - \Omega \tau $,
in terms of which the metric (\ref{eq:metric}) becomes, for $n=3$,
\begin{subequations}
\label{eq:rotmetric}
    \begin{multline}
     {\mathrm {d}}s^{2}
     =  a^{-2} \sec ^{2}\rho  \, \bigg[ - \left( 1 - \Omega ^{2} \sin ^{2} \rho \right) {\mathrm {d}}\tau ^{2} + 2 \Omega \sin ^{2} \rho \,   {\mathrm {d}} \tau \, {\mathrm {d}}\varphi 
     \\
     + {\mathrm {d}}\rho ^{2} + \sin ^{2} \rho \,  {\mathrm {d}} \varphi  ^{2}  \bigg]  ,
     \end{multline}
     while for $n=4$ we have
\begin{multline}
{\mathrm {d}}s^{2}
     =
      a^{-2} \sec ^{2}\rho \,  \bigg[ - \left( 1 - \Omega ^{2} \sin ^{2} \rho \sin ^{2} \theta \right) {\mathrm {d}}\tau ^{2}  + {\mathrm {d}}\rho ^{2}
      \\
     + 2\Omega \sin ^{2} \rho \sin ^{2} \theta  \, {\mathrm {d}}\tau \, {\mathrm {d}}\varphi 
      + \sin ^{2} \rho \,  \left\{ {\mathrm {d}}\theta ^{2}
      + \sin ^{2} \theta \, {\mathrm {d}} \varphi ^{2}  \right\} \bigg]  .
\end{multline}
\end{subequations}
Particles at constant $(\rho, \varphi )$ ($n=3$) or constant $(\rho, \theta, \varphi )$ ($n=4$) have a speed which is greater than the speed of light if the metric component $g_{\tau \tau }>0$.
From the line elements (\ref{eq:rotmetric}), this cannot happen if $|\Omega |<1$.
For the rest of this paper, we therefore assume this inequality holds. 

In our RKT analysis in the next section, we will require an orthonormal dreibein/vierbein of basis vectors $e_{(a)}$ for the metric (\ref{eq:rotmetric}), where drei/vierbein indices are enclosed in round brackets.
These have a corresponding basis of one-forms $\omega ^{(a)}$ given by, for both $n=3$ and $n=4$,
\begin{subequations}
\label{eq:oneforms}
\begin{equation}
    \omega ^{(\rho )} = a^{-1} \sec \rho \, {\mathrm {d}}\rho ,
\end{equation}
while for $n=3$ we also have
    \begin{align}
    \omega ^{(t,3)}  & = a^{-1}\sec \rho  \, \left( \Gamma ^{-1} \, {\mathrm {d}}\tau -  \Omega \Gamma \sin ^{2}\rho  \, {\mathrm {d}}\varphi  \right),
    \\
    \omega ^{(\varphi ,3)} & = a^{-1}\Gamma \tan \rho \, {\mathrm {d}}\varphi , 
    \end{align}
    and, for $n=4$,
        \begin{align}
    \omega ^{(t,4)}  & = a^{-1}\sec \rho  \, \left( \Gamma ^{-1} \, {\mathrm {d}}\tau -  \Omega \Gamma \sin ^{2}\rho \sin ^{2}\theta  \, {\mathrm {d}}\varphi  \right),
    \\
    \omega ^{(\theta )} & = a^{-1}\tan \rho \, {\mathrm {d}}\theta ,
    \\
    \omega ^{(\varphi ,4)} & = a^{-1}\Gamma \tan \rho  \sin \theta \, {\mathrm {d}}\varphi , 
    \end{align}
\end{subequations}
where we have defined
\begin{equation}
\Gamma = \begin{cases}
\left( 1-\Omega^2\sin^2\rho\right) ^{-\frac{1}{2}}, & n=3,
     \\
\left( 1-\Omega^2\sin^2\rho\sin^2\theta\right) ^{-\frac{1}{2}}, & n=4.
\end{cases}
\label{eq:Gamma}
\end{equation}

\section{Relativistic kinetic theory analysis}
\label{sec:RKT}

In this section we use RKT to find the SET (which we will refer to as the RKT-SET from now on) for a rigidly-rotating thermal gas of massless bosonic particles in AdS. 
We model the gas of classical particles as a fluid described by the following distribution function in $n$ dimensions \cite{Boisseau:1989}
\begin{equation}
    f = \frac{1}{(2\pi)^{n-1}}\frac{1}{\exp \left[ {\widetilde {\beta }} u_{(a)} p^{(a)} \right]-1},
    \label{eq:fRKT}
\end{equation}
where $u^{(a)}$ and $p^{(a)}$ are the $n$-velocity and $n$-momentum of the fluid particles, ${\widetilde {\beta }}$ is the (local) inverse temperature and we have assumed that the chemical potential vanishes. 
We assume that the fluid is at rest relative to the frame of vectors $e_{(a)}$, so that we are considering a rigidly-rotating distribution. 
We then have $u_{(a)}p^{(a)}=p^{(0)}$, which simplifies both the distribution function (\ref{eq:fRKT}), and the integrals required to find the RKT-SET, which is given by
\begin{equation}
    {}^{n}T_{\mathrm {RKT}}^{(a)(b)} = \int \frac{{\mathrm {d}}^{n-1}{\mathbf {p}}}{p^{(0)}} f \, p^{(a)}p^{(b)},
    \label{eq:RKTSET}
\end{equation}
where the integral is taken over the spatial components of the $n$-momentum.
Since we are considering massless particles, we have $p^{(0)}=|{\mathbf {p}}|$, which also simplifies the integrals in (\ref{eq:RKTSET}).

In a curved space-time, the local inverse temperature ${\widetilde {\beta }}$ is not a constant, but given by \cite{Tolman:1930zza,Tolman:1930ona} 
\begin{equation}
\label{eq:loctemp}
    {\widetilde {\beta }} = \beta {\sqrt {-g_{\tau \tau }}} 
    =   a^{-1} \beta \Gamma ^{-1} \sec \rho   ,
\end{equation}
where $\beta $ is the inverse temperature at the origin $\rho =0$, and we have used the line element (\ref{eq:rotmetric}) and the definition (\ref{eq:Gamma}).
The local inverse temperature (\ref{eq:loctemp}) depends on the radial coordinate $\rho $ when $n=3$ and on both $\rho $ and the polar angle $\theta $ when $n=4$.
From (\ref{eq:loctemp}), we have that ${\widetilde {\beta }}$ diverges as $\rho \to \pi /2$, and hence the local temperature ${\widetilde {\beta }}^{-1}$ vanishes on the AdS boundary, as is the case for nonrotating states \cite{Ambrus:2018olh}. 

The momentum-space integrals in (\ref{eq:RKTSET}) are straightforward to compute in both three and four dimensions. 
We find the drei/vierbein components of the RKT-SET to be
\begin{equation}
    {}^{n}T^{(a)(b)}_{\mathrm {RKT}} = 
    \begin{cases}
        {\displaystyle {\frac{\zeta (3)}{2\pi {\widetilde {\beta }}^{3}} {\mathrm {Diag}} \, \{ 2, 1, 1 \} }} , & n=3,
        \\ \\
        {\displaystyle {\frac{\pi ^{2}}{90{\widetilde {\beta }}^{4}} {\mathrm {Diag}} \, \{ 3, 1, 1, 1 \} }} , & n=4,
    \end{cases}
    \label{eq:RKT-SET}
\end{equation}
where $\zeta (3) \approx 1.20206$ is the Riemann zeta function. 
It is straightforward to check that the RKT-SET (\ref{eq:RKT-SET}) is traceless and conserved.
In the zero-temperature limit $\beta \rightarrow \infty $, the RKT-SET (\ref{eq:RKT-SET}) vanishes. Thus the RKT approximation does not incorporate 
vacuum energy effects \cite{Ambrus:2017vlf,Ambrus:2018olh}. For this reason, in the following sections, we will compare the RKT-SET (\ref{eq:RKT-SET}) with the difference in RSET expectation values between rigidly-rotating thermal and vacuum states.

\section{Quantum field theory analysis}
\label{sec:QFT}

We consider a massless, conformally coupled scalar field $\Phi $ satisfying the field equation
\begin{equation}
\left( \nabla _{\mu} \nabla ^{\mu } - \xi R \right) \Phi =0,
\end{equation}
where $\xi $ is the coupling constant and $R$ the Ricci scalar curvature, which equals $-6a^{2}$ when $n=3$ and $-12a^{2}$ when $n=4$.
Considering conformal coupling, we have $\xi = 1/8$ for $n=3$ and $\xi = 1/6$ for $n=4$.
In the absence of a speed-of-light surface, the rigidly-rotating vacuum state is the same as the nonrotating AdS vacuum \cite{Kent:2014wda,Kent:2013ouo}, and the RSET for the massless, 
conformally-coupled scalar field in the vacuum state is \cite{Kent:2014nya}
\begin{equation}
\langle {\hat {T}}_{\mu \nu } \rangle _{0} = 
\begin{cases}
    0, & n=3, \\
    {\displaystyle {
- \frac{a^{4}}{960 \pi ^{2}} g_{\mu \nu } , 
    }} & n =4.
\end{cases}
\label{eq:vacRSET}
\end{equation}
Since the difference in expectation values between two quantum states does not require renormalization, to find the RSET for a rigidly-rotating thermal state we consider the difference
\begin{equation}
    \Delta {\hat {T}}_{\mu \nu } = \langle {\hat {T}}_{\mu \nu } \rangle _{\beta } - \langle {\hat {T}}_{\mu \nu } \rangle _{0}
    \label{eq:RSETdiff}
 \end{equation}
between the RSET in a rigidly-rotating thermal state at inverse temperature $\beta $ and the vacuum RSET (\ref{eq:vacRSET}). 
This difference in expectation values is constructed as follows.

We start with the vacuum Green function ${}^{n}G_{0}(x,x')$ for the scalar field (that is, $-{\mathrm {i}}$ multiplied by the Feynman propagator), which takes the form 
\begin{subequations}
\label{eq:G0}
\begin{align}
    {}^{3}G_{0}(x,x') & = \frac{a}{4{\sqrt {2}}\pi } \left[ \frac{1}{\sqrt {Z-1}} -\frac{1}{\sqrt{Z+1}}\right] ,
    \\
    {}^{4}G_{0}(x,x') & = \frac{a^{2}}{8\pi ^{2}} \left[ \frac{1}{Z-1} - \frac{1}{Z+1}\right] ,
\end{align}
\end{subequations}
in three and four dimensions, respectively, where
\begin{equation}
    Z(x,x') = \sec \rho \sec \rho ' \cos \left(\tau - \tau ' \right) -\tan \rho \tan \rho ' \cos \gamma _{n} ,
\end{equation}
and, in rotating coordinates, 
\begin{subequations}
   \begin{align}
        \cos \gamma _{3}&  = 
            \cos \left( \varphi - \varphi ' + \Omega \left[ \tau - \tau' \right] \right), 
            \\
           \cos \gamma _{4}&  =  \cos \theta \cos \theta ' + \sin \theta \sin \theta ' \cos \left(  \varphi - \varphi ' + \Omega \left[ \tau - \tau' \right] \right) .
    \end{align}
\end{subequations}
Our purpose in this paper is to compare the results for the RKT-SET (\ref{eq:RKT-SET}) with the QFT-RSET which we derive in this section.
In RKT, the quantum scalar field is modelled as a thermal gas of classical particles and thus there can be no flux of energy through the AdS boundary at $\rho = \pi/2$.
For a scalar field on AdS, it is necessary to apply boundary conditions to have a well-defined quantum field theory \cite{Avis:1977yn}. 
In order for there to be no flux of energy through the AdS boundary, the scalar field must satisfy either Dirichlet or Neumann boundary conditions
\cite{deOliveira:2022gwd}.  
Comparing the results of \cite{Ambrus:2018olh,Morley:2023exv} for a massless, conformally-coupled, quantum scalar field in a nonrotating thermal state in four-dimensional AdS, it can be seen that it is the RSET when Dirichlet boundary conditions are applied which is better approximated by the corresponding RKT quantity, rather than that with Neumann boundary conditions applied.
We have therefore applied Dirichlet boundary conditions to the quantum scalar field in the vacuum Green function (\ref{eq:G0}). 

The Green function for a rigidly-rotating thermal state with inverse temperature $\beta $ is then given by the Matsubara sum
\begin{equation}
    {}^{n}G_{\beta }(x,x')= \sum _{j=-\infty }^{\infty } {}^{n}G_{0}(x_{\beta }, x') ,
    \label{eq:Gthermal}
\end{equation}
where 
\begin{equation}
    x_{\beta } = \begin{cases}
        (\tau + {\mathrm {i}}j\beta , \rho ,\varphi ), & n=3,
        \\
        (\tau + {\mathrm {i}}j\beta , \rho ,\theta , \varphi ), & n=4.
    \end{cases}
\end{equation}
The difference $\Delta G (x,x')={}^{n}G_{\beta }(x,x')-{}^{n}G_{0}(x,x')$ between the thermal and vacuum Green functions corresponds to simply removing the $j=0$ term in the sum in (\ref{eq:Gthermal}). 

The difference in RSETs (\ref{eq:RSETdiff}) is built up from the coincidence limit of $\Delta G(x,x')$ (which is finite) and its derivatives \cite{Decanini:2005eg}:
\begin{equation}
     \Delta {\hat {T}}_{\mu \nu } = -w_{\mu \nu } + \frac{\left( 1 - 2\xi \right)}{2} \nabla _{\mu } \nabla _{\nu } w
     + \frac{\left( 4\xi - 1 \right)}{4} g_{\mu \nu } \nabla _{\lambda }\nabla ^{\lambda } w 
     +\frac{\xi  R}{n} g_{\mu \nu }w ,
\end{equation}
where
\begin{equation}
    w = \lim _{x'\to x} \left[\Delta G (x,x') \right] , \quad w_{\mu \nu } = \lim_{x'\to x} \left[ \nabla _{\mu }\nabla _{\nu } \left\{  \Delta G(x,x') \right\}  \right] ,
\end{equation} 
and we have used the fact that the Ricci tensor on AdS is $R_{\mu \nu } = n^{-1}Rg_{\mu \nu }$.

After some lengthy algebra, we find that $\Delta {\hat {T}}_{\mu \nu }$  takes the following form in three dimensions:
\begin{equation}
         {}^{3}\Delta {\hat {T}}_{\mu \nu }= 
         \frac{a}{32\sqrt{2}\pi}\sum_{j\neq 0} \varpi_{\mu \nu }[{\mathcal{A}}_{+}] - \varpi_{\mu \nu }[{\mathcal{A}}_{-}] ,
\end{equation}
where  we have defined 
\begin{equation}
     \mathcal{A}_\pm = \pm 1 + \cosh (j \beta ) \sec ^2 \rho -\cosh (j \beta  \Omega ) \tan ^2 \rho ,
\end{equation}
and the nonzero quantities $\varpi_{\bullet \bullet}[{\mathcal{A}}_{\pm}]$ are
\begin{subequations}
    \begin{align}
         \varpi_{\tau\tau}[\mathcal{A}_\pm] & =
         -\frac{1}{(\mathcal{A}_\pm)^{\frac{5}{2}}}\left\{ {}_2\mathcal{B}_{\tau\tau}(\mathcal{A}_\pm)^2+ {}_1\mathcal{B}_{\tau\tau}(\mathcal{A}_\pm) + {}_{0}\mathcal{B}_{\tau\tau} \right\} ,
         \\
         \varpi_{\tau\varphi }[\mathcal{A}_\pm] & = 
         \frac{6}{(\mathcal{A}_\pm)^{\frac{5}{2}}}\sec^2\rho\tan^2 \rho\sinh(j\beta)\sinh(j\beta\Omega) 
         \nonumber \\ &  \qquad +
         \Omega\,\varpi_{\varphi \varphi }[\mathcal{A}_\pm] ,
         \\
          \varpi_{\rho\rho}[\mathcal{A}_\pm] & =\frac{\sec^2 \rho}{(\mathcal{A}_\pm)^{\frac{3}{2}}}\left\{ 2(\mathcal{A}_\pm) - {}_{0}\mathcal{B}_{\rho\rho} \right\} ,
          \\
          \varpi_{\varphi \varphi }[\mathcal{A}_\pm] &  =\dfrac{\tan^2 \rho }{(\mathcal{A}_\pm)^{\frac{5}{2}}}\left\{ 2(\mathcal{A}_\pm)^2+{}_{1}\mathcal{B}_{\varphi \varphi }(\mathcal{A}_\pm) + {}_{0}\mathcal{B}_{\varphi \varphi } \right\} ,
    \end{align}
\end{subequations}
with coefficients
\begin{subequations}
    \begin{align}
         {}_2\mathcal{B}_{\tau\tau} & =  2 \left[ \Omega^2-\left(\Omega^2-1\right)\sec^2\rho \right] ,
         \\
    {}_1\mathcal{B}_{\tau\tau} & =  \cosh(j\beta)\left[ 2\Omega^2 - (7\Omega^2+1)\sec^2\rho+5(\Omega^2-1)\sec^4\rho \right] 
    \nonumber \\ & \qquad  
   - \cosh(j\beta\Omega)\left[ 6\Omega^2 - (11\Omega^2-3)\sec^2\rho
    \right. \nonumber \\ & \left. \qquad \qquad +5(\Omega^2-1)\sec^4\rho \right] ,
    \\
    {}_0\mathcal{B}_{\tau\tau} & =  3\tan^2 \rho \sec^2\rho \left[\Omega^2- \left(\Omega^2-1\right)\sec^2\rho\right] 
    \nonumber \\ & \qquad \qquad  \times \left[ \cosh(j\beta)-\cosh(j\beta\Omega)\right]^2 
     \nonumber \\ & \qquad 
     +6\left[ \sinh(j\beta)\sec^2\rho - \Omega\sinh(j\beta\Omega)\tan^2\rho \right]^2 ,
     \\
     {}_{0}\mathcal{B}_{\rho\rho} & = \cosh(j\beta)\left[ 1+2\sec^2\rho  \right] +\cosh(j\beta\Omega)\left[ 3-2\sec^2\rho \right] ,
     \\
      {}_1\mathcal{B}_{\varphi\varphi} & = \cosh(j\beta)\left[ 2-5\sec^2\rho\right]
     - \cosh(j\beta\Omega)\left[ 6 - 5\sec^2\rho\right] ,
      \\
    {}_0\mathcal{B}_{\varphi\varphi } &=  3\tan^2\rho\, \left\{ \sec^2\rho \left[ \cosh(j\beta)-\cosh(j\beta\Omega)\right]^2 
   \right. \nonumber \\ & \left. \qquad \qquad
    -2\sinh^2(j\beta\Omega)  \right\}.
    \end{align}
\end{subequations}
The corresponding expressions in four dimensions are too lengthy to reproduce here, but can be found in a {\tt {Mathematica}} notebook provided in the Supplementary Material \cite{Tmunu4d}.

\begin{figure*}
    \centering
    \includegraphics[width=0.45\linewidth]{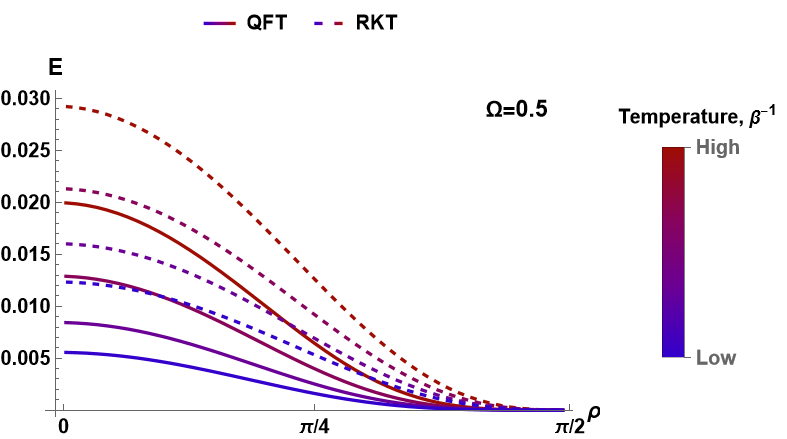}
    \includegraphics[width=0.45\linewidth]{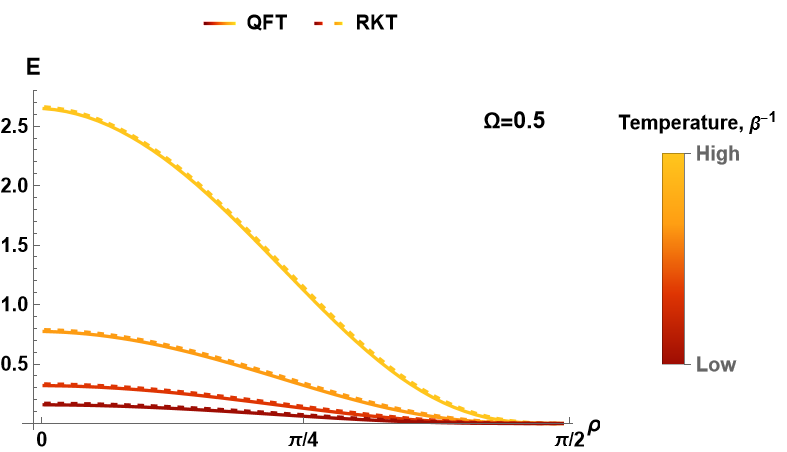}
    \\
    \includegraphics[width=0.45\linewidth]{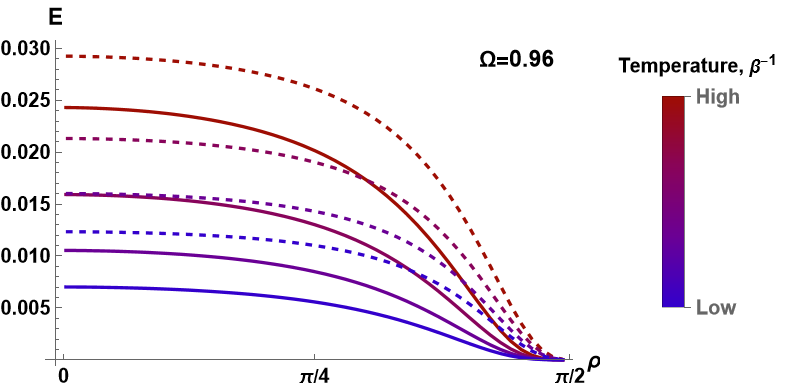}
    \includegraphics[width=0.45\linewidth]{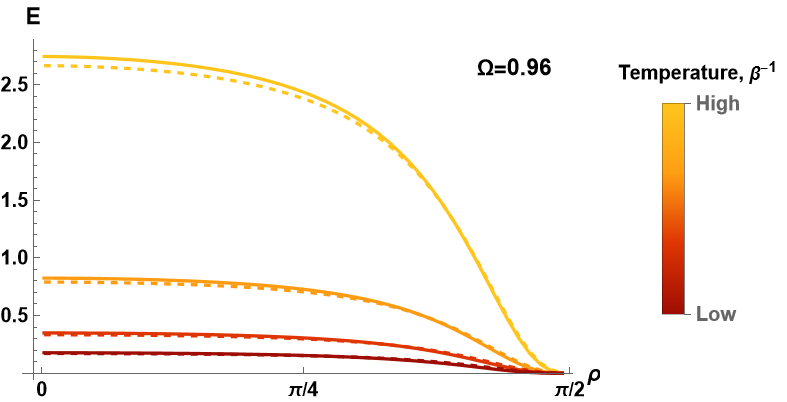}
    \caption{Energy density $E$ for a rigidly-rotating thermal state in three-dimensional AdS, for angular speed $\Omega = 0.5$ (top row) and $\Omega = 0.96$ (bottom row). 
    In the left-hand-plots, the inverse temperature $\beta $ takes the values $\beta \in \{ 3\pi/4, 5\pi /6, 11\pi /12, \pi \}$, while in the right-hand-plots we have $\beta \in \{ \pi/6, \pi /4, \pi /3, 5\pi /12 \}$.  Solid lines are the results for the QFT-RSET, while dotted lines are results from the RKT-SET. }
    \label{fig:3dEnergy}
\end{figure*}

For comparison with the RKT-SET, we convert the space-time components $\Delta {\hat {T}}^{\mu \nu }$ to frame components using the one-forms (\ref{eq:oneforms}):
\begin{equation}
    {}^{n}T^{(a)(b)}_{\mathrm {QFT}} = \omega ^{(a)}_{\mu }\omega ^{(b)}_{\nu} \Delta {\hat {T}}^{\mu \nu },
    \label{eq:QFTRSET}
\end{equation}
and from now on we will refer to (\ref{eq:QFTRSET}) as the QFT-RSET.
It should be emphasized that, for comparison with our RKT results, we are interested in the difference in QFT expectation values between the rigidly-rotating thermal and vacuum states, and therefore (\ref{eq:QFTRSET}) will be a traceless tensor in both three and four dimensions. If one were interested in the thermal expectation value of the RSET in QFT (as derived, for example, using Hadamard renormalization), it would be necessary to add the anomalous trace contribution (\ref{eq:vacRSET}) to the difference (\ref{eq:RSETdiff}).

\section{Comparing the QFT-RSET and the RKT-SET}
\label{sec:QC}

\begin{figure*}
    \centering
    \includegraphics[width=0.45\linewidth]{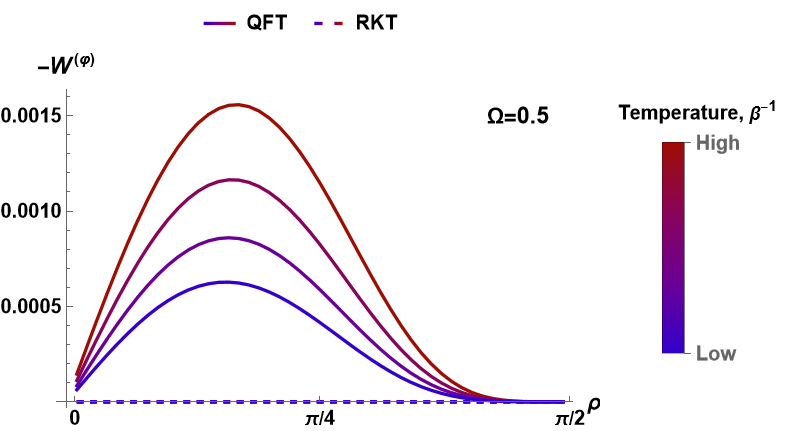}
    \includegraphics[width=0.45\linewidth]{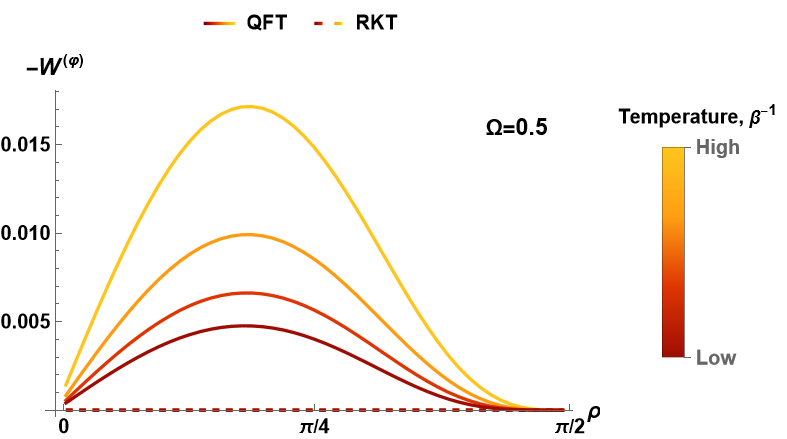}
    \\
    \includegraphics[width=0.45\linewidth]{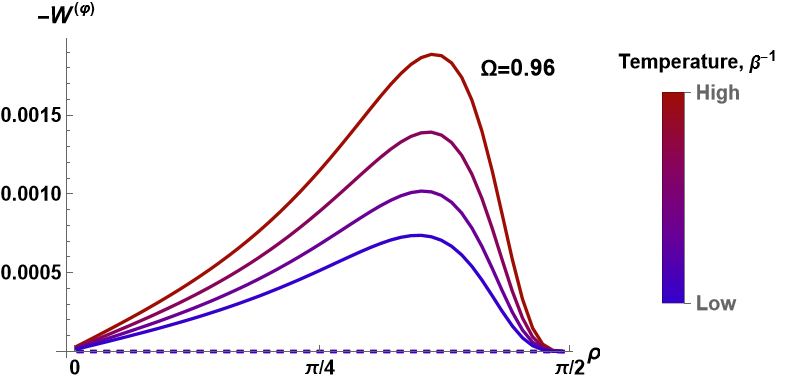}
    \includegraphics[width=0.45\linewidth]{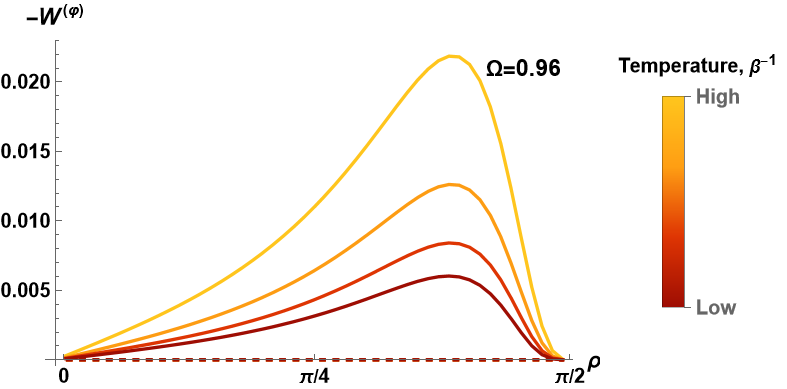}
    \caption{Heat flux $-W^{(\varphi )}$ for a rigidly-rotating thermal state in three-dimensional AdS, for angular speed $\Omega = 0.5$ (top row) and $\Omega = 0.96$ (bottom row). 
    In the left-hand-plots, the inverse temperature $\beta $ takes the values $\beta \in \{ 3\pi/4, 5\pi /6, 11\pi /12, \pi \}$, while in the right-hand-plots we have $\beta \in \{ 
\pi/6, \pi /4, \pi /3, 5\pi /12 \}$.  Solid lines are the results for the QFT-RSET, while dotted lines are results from the RKT-SET (which are identically zero). 
}
    \label{fig:3dHeatFlux}
\end{figure*}

\begin{figure*}
    \centering
    \includegraphics[width=0.45\linewidth]{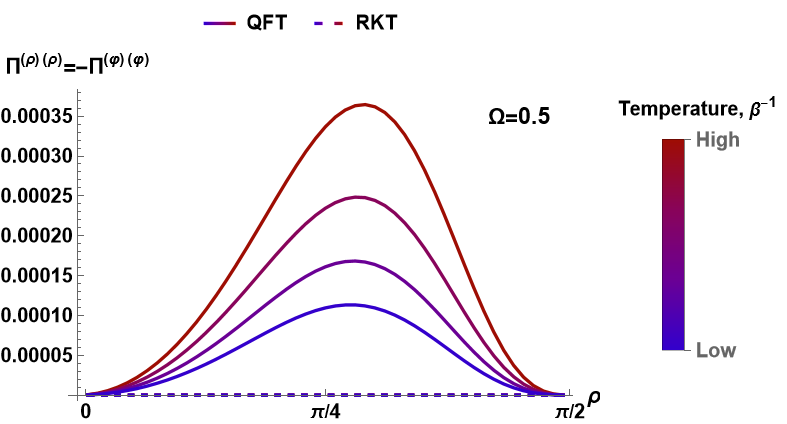}
    \includegraphics[width=0.45\linewidth]{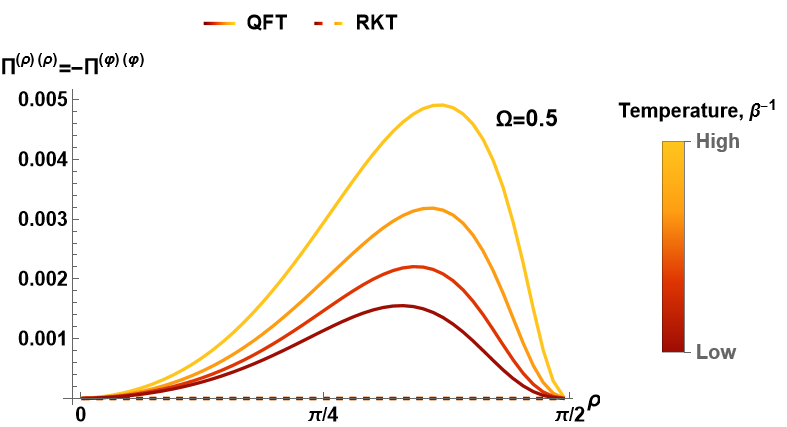}
    \\
    \includegraphics[width=0.45\linewidth]{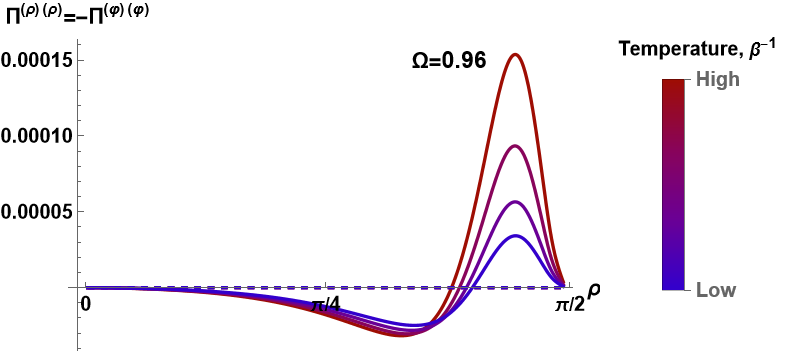}
    \includegraphics[width=0.45\linewidth]{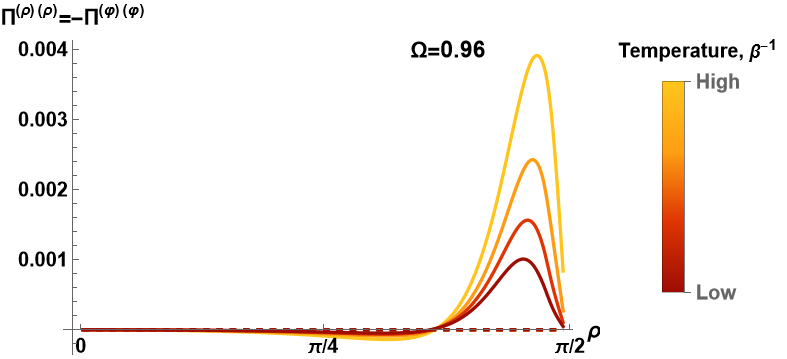}
    \caption{Pressure deviator $\Pi ^{(\rho )(\rho )}=-\Pi ^{(\varphi)(\varphi )}$ for a rigidly-rotating thermal state in three-dimensional AdS, for angular speed $\Omega = 0.5$ (top row) and $\Omega = 0.96$ (bottom row). 
    In the left-hand-plots, the inverse temperature $\beta $ takes the values $\beta \in \{ 3\pi/4, 5\pi /6, 11\pi /12, \pi \}$, while in the right-hand-plots we have $\beta \in \{  \pi/6, \pi /4, \pi /3, 5\pi /12 \}$.  Solid lines are the results for the QFT-RSET, while dotted lines are results from the RKT-SET (which are identically zero). 
    }
    \label{fig:3dPresDev}
\end{figure*}

In order to compare the SETs arising from RKT and QFT, we employ the thermometer frame decomposition \cite{Van:2013sma,Becattini:2014yxa}, writing, with $\eta ^{(a)(b)}$ the Minkowski metric,
\begin{multline}
    {}^{n}T^{(a)(b)}_{\mathrm{RKT/QFT}} = \left( E + P \right) u^{(a)}u^{(b)} + P \eta ^{(a)(b)} + u^{(a)}W^{(b)} + W^{(a)}u^{(b)}
    \\ +\Pi ^{(a)(b)},
\end{multline}
where $E$ is the energy density, $P$ the isotropic pressure, $W^{(a)}$ the heat flux and $\Pi ^{(a)(b)}$ the anisotropic stress.
These quantities can be computed from the components of the SET in the dreibein/veirbein frame as follows \cite{Ambrus:2021eod}:
\begin{subequations}
\label{eq:quantities}
\begin{align}
    E & = u^{(a)}u^{(b)}T_{(a)(b)},
    \\
    P & = \frac{1}{n-1}\Delta ^{(a)(b)} T_{(a)(b)} ,
    \\
    W^{(a)} & = -\Delta ^{(a)(b)}u^{(c)}T_{(b)(c)},
    \\
    \Pi ^{(a)(b)} & = \left[ \Delta ^{(a)(c)} \Delta ^{(b)(d)} - \frac{1}{n-1}\Delta ^{(a)(b)}\Delta ^{(c)(d)}\right] T_{(c)(d)},
\end{align}
where $\Delta ^{(a)(b)} = u^{(a)}u^{(b)} + \eta ^{(a)(b)}$ is the projector. 
\end{subequations}
In RKT, the heat flux $W^{(a)}$ and anisotropic stress $\Pi ^{(a)(b)}$ vanish identically, while $E=(n-1)P$ since we are considering massless particles.
The QFT-RSET (\ref{eq:QFTRSET}) is also traceless and has $E=(n-1)P$. 
Hence we may quantify quantum effects by considering the nonzero components of $W^{(a)}$ and $\Pi ^{(a)(b)}$, which are $W^{(\varphi )}$ in both three and four dimensions, and ${}^{3}\Pi ^{(\rho )(\rho )}=-{}^{3}\Pi ^{(\varphi )(\varphi )}$ for $n=3$, while for $n=4$, the nonzero components of the anisotropic stress are ${}^{4}\Pi ^{(\rho )(\rho )}$, ${}^{4}\Pi ^{(\rho )(\theta )}$, ${}^{4}\Pi ^{(\theta  )(\theta  )}$  and ${}^{4}\Pi ^{(\varphi )(\varphi )}$, satisfying ${}^{4}\Pi ^{(\rho )(\rho )} + {}^{4}\Pi ^{(\theta )(\theta )} + {}^{4}\Pi ^{(\varphi )(\varphi )}=0$.
When $\Omega = 0$, it is the case that ${}^{4}\Pi ^{(\theta )(\theta )}={}^{4}\Pi ^{(\varphi )(\varphi )}$, but this does not hold for rigidly-rotating states.

\subsection{Three dimensions}
\label{sec:compare3}

We begin, in Fig.~\ref{fig:3dEnergy}, by comparing the energy density $E$ in three dimensions for the RKT-SET and QFT-RSET.
For all values of the inverse temperature $\beta $ and angular speed $\Omega $ shown, the energy density has a maximum at the origin and monotonically decreases to zero on the space-time boundary. 
The rate at which $E$ decays to zero is lower for higher $\Omega $ and higher temperatures.
At high temperatures (right-hand-plots), the RKT-SET is an excellent approximation to the QFT-RSET, even at high angular speeds (bottom row).  
In contrast, at low temperatures (left-hand-plots), quantum effects are significant, with the RKT-SET giving an energy density which is considerably larger than that arising in QFT.
The pressure $P=E/2$ in both QFT and RKT and therefore has the same properties as the energy density $E$.
 
We now study the quantities $W^{(b)}$ and $\Pi ^{(a)(b)}$, which vanish identically in the RKT approximation. 
The heat flux $W^{(b)}$ has a single nonzero component, 
$W^{(\varphi )}$, which is shown in Fig.~\ref{fig:3dHeatFlux}.
This quantity vanishes both at the origin $\rho =0$ and on the boundary $\rho \rightarrow \pi /2$. 
Elsewhere, the magnitude of $W^{(\varphi )}$ is at least an order of magnitude smaller than that of the energy density $E$.
For all values of $\beta  $ and $\Omega $ studied, the heat flux is negative, indicating that, in the rigidly-rotating frame, the heat flux is in the negative $\varphi $ direction, that is, opposite to the direction of rotation. 
This implies that the quantum radiation is rotating less quickly than the angular speed $\Omega $ corresponding to the rigid rotation.
Finally, we observe that the location of the minimum of the heat flux moves closer to the AdS boundary as the angular speed $\Omega $ increases.

The single nonzero component of the pressure deviator, $\Pi _{\beta }^{(\rho )( \rho )}=-\Pi _{\beta } ^{(\varphi )(\varphi )}$ is shown in Fig.~\ref{fig:3dPresDev}.
Like the heat flux, this vanishes at the origin and AdS boundary.
Its magnitude is  roughly an order of magnitude smaller than that of the heat flux.
For the smaller rotation speed, $\Pi _{\beta }^{(\rho )(\rho )}$ is positive throughout the space-time, with its maximum slightly closer to the AdS boundary at high temperatures. 
When $\Omega = 0.96$, we observe rather different behaviour, with $\Pi _{\beta }^{(\rho )( \rho)}$ negative close to the origin, and positive close to the AdS boundary, with a sharp peak which is even closer to the boundary at high temperatures.  

\begin{figure*}
    \centering
    \includegraphics[width=0.45\linewidth]{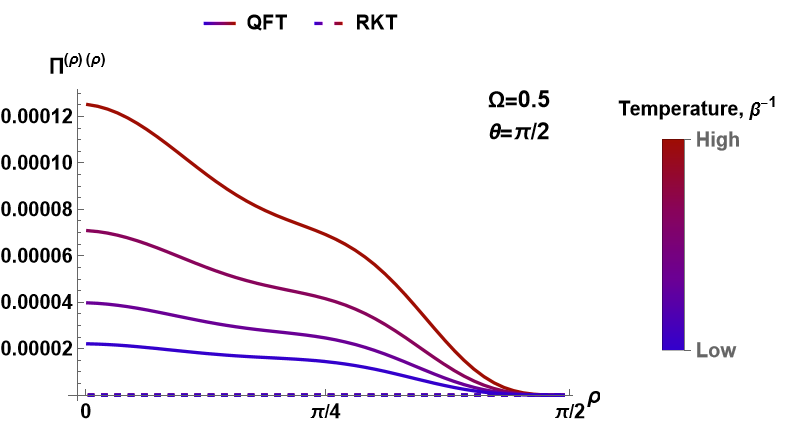}
    \includegraphics[width=0.45\linewidth]{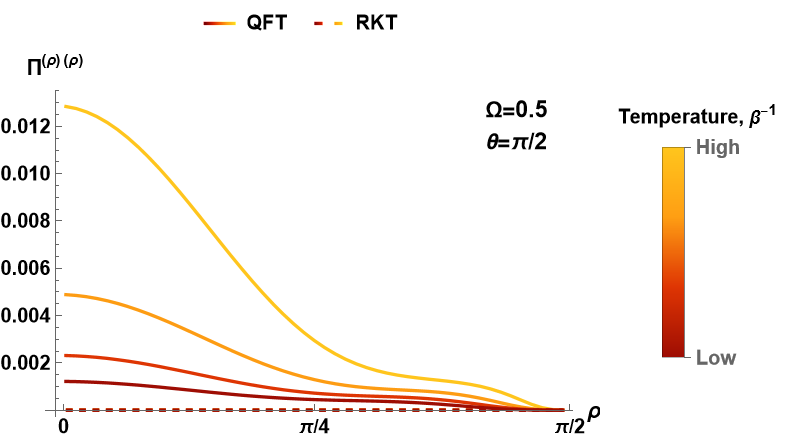}
    \\
    \includegraphics[width=0.45\linewidth]{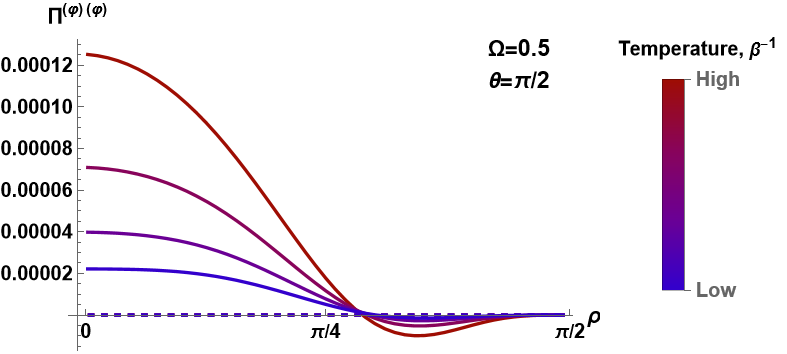}
    \includegraphics[width=0.45\linewidth]{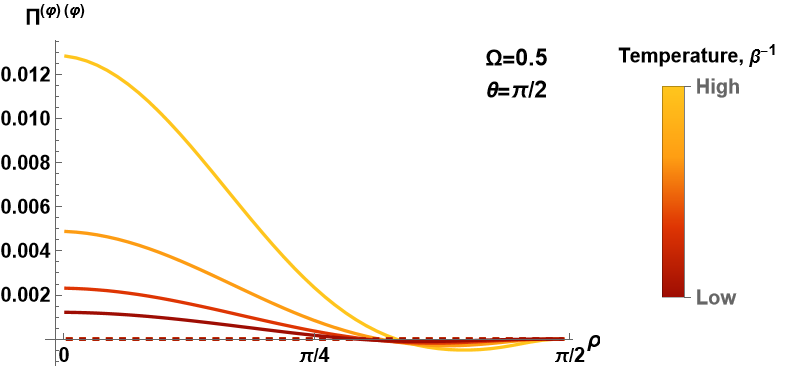}
    \caption{Pressure deviator components $\Pi ^{(\rho )(\rho )}$ (top row) and $\Pi ^{(\varphi )(\varphi )}$ (bottom row) for a rigidly-rotating thermal state in the equatorial plane ($\theta = \pi/2$) of four-dimensional AdS, for angular speed $\Omega = 0.5$. 
    In the left-hand-plots, the inverse temperature $\beta $ takes the values $\beta \in \{ 3\pi/4, 5\pi /6, 11\pi /12, \pi \}$, while in the right-hand-plots we have $\beta \in \{  \pi/6, \pi /4, \pi /3, 5\pi /12 \}$. Solid lines are the results for the QFT-RSET, while dotted lines are results from the RKT-SET (which are identically zero). 
    }
    \label{fig:4dPresDevRhoPhi}
\end{figure*}

\begin{figure*}
   \centering
    \includegraphics[width=0.23\linewidth]{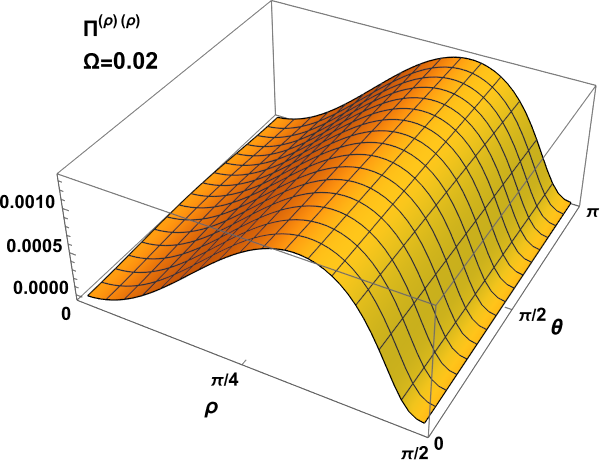} \quad
     \includegraphics[width=0.23\linewidth]{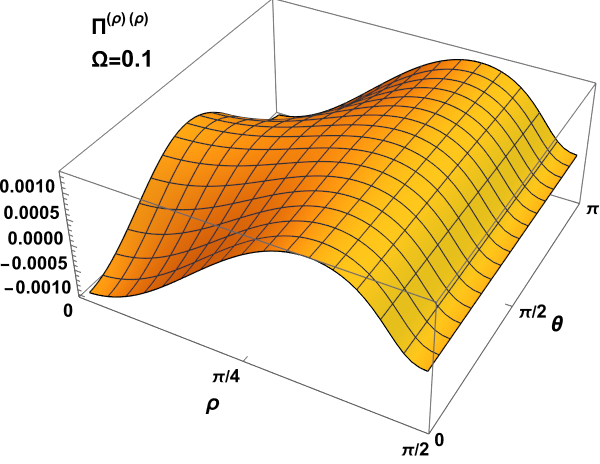} \quad
        \includegraphics[width=0.23\linewidth]{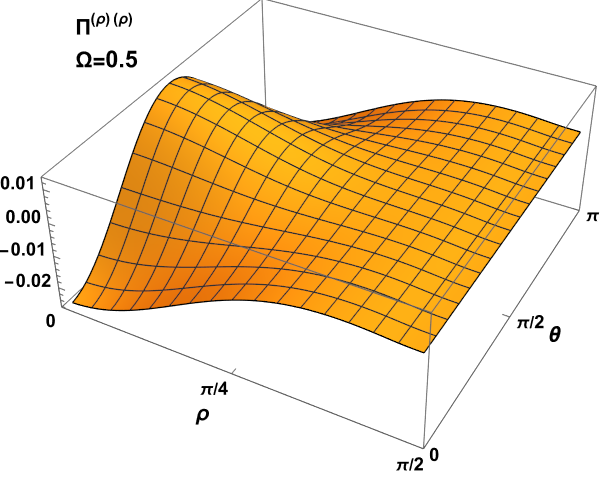} \quad
    \includegraphics[width=0.23\linewidth]{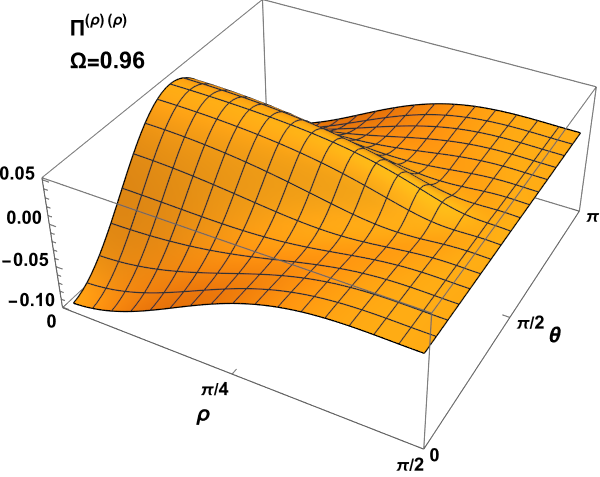} 
    \medskip \\
    \includegraphics[width=0.23\linewidth]{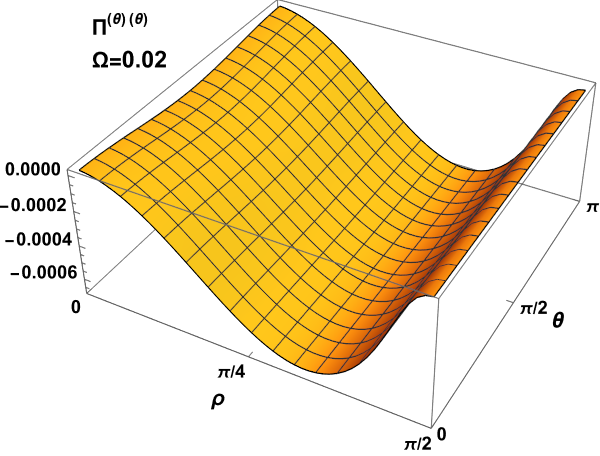} \quad
    \includegraphics[width=0.23\linewidth]{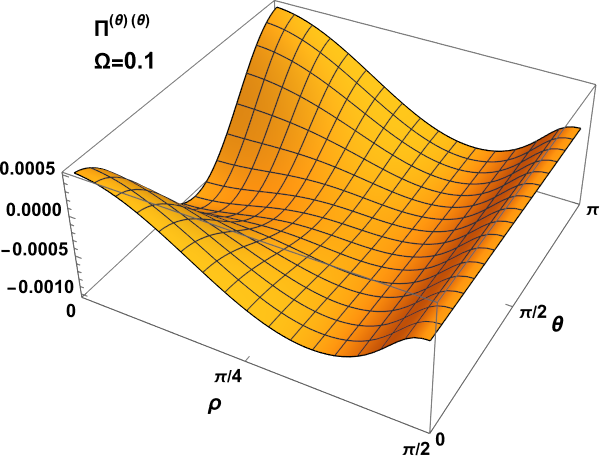} \quad
    \includegraphics[width=0.23\linewidth]{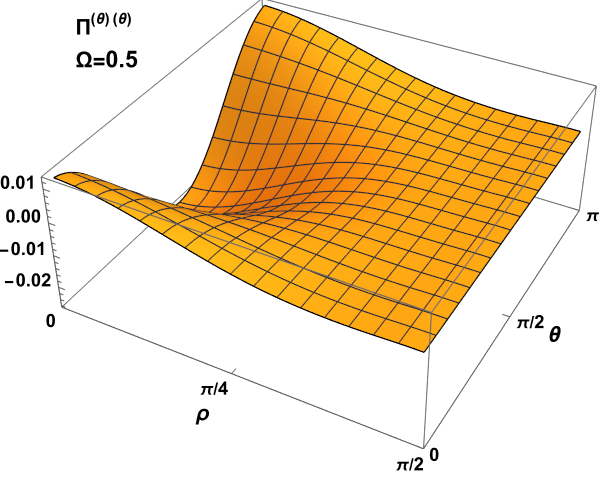} \quad
    \includegraphics[width=0.23\linewidth]{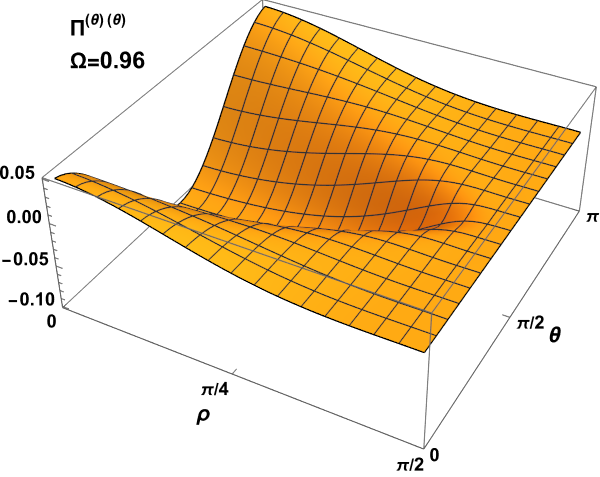}
\medskip \\
    \includegraphics[width=0.23\linewidth]{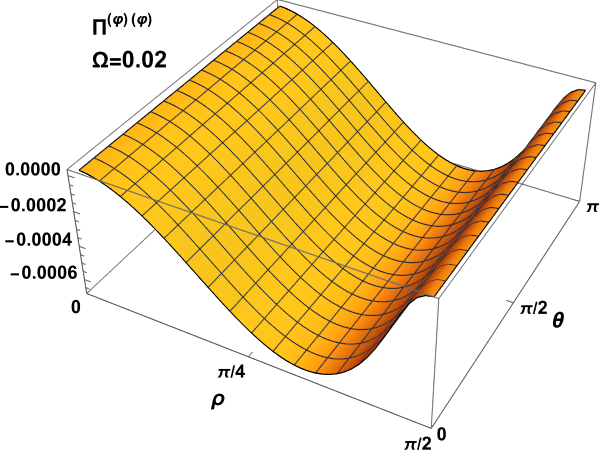} \quad
    \includegraphics[width=0.23\linewidth]{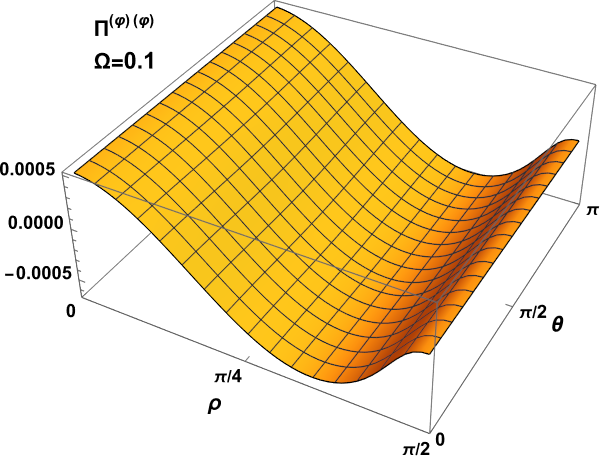} \quad
    \includegraphics[width=0.23\linewidth]{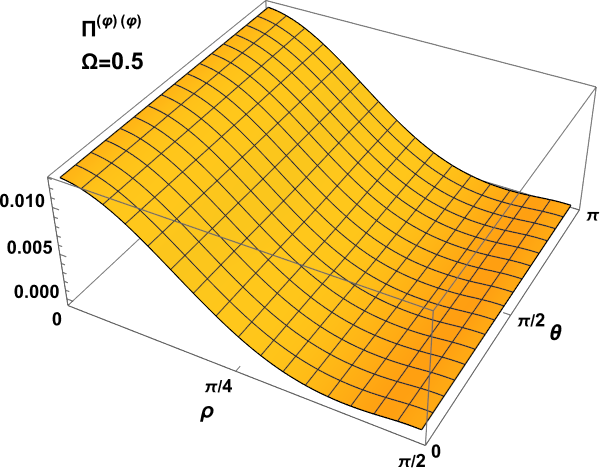} \quad
    \includegraphics[width=0.23\linewidth]{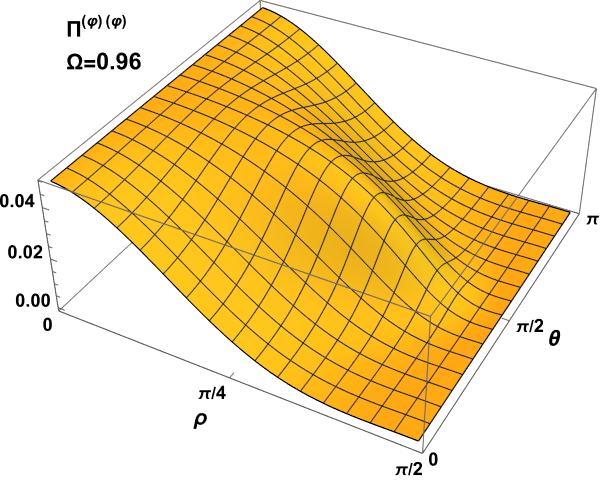}
\caption{Components of the pressure deviator $\Pi ^{(\rho )(\rho )}$ (top row), $\Pi ^{(\theta )(\theta )}$ (middle row) and $\Pi ^{(\varphi )(\varphi )}$ (bottom row) for a rigidly-rotating thermal state on four-dimensional AdS, for angular speeds $\Omega = 0.02$ (left column), $\Omega = 0.1$ (left-centre column), $\Omega = 0.5$ (right-centre column) and $\Omega = 0.96$ (right column). The inverse temperature is $\beta = \pi /6$. 
}
\label{fig:4dPresDevtheta}
\end{figure*}

\subsection{Four dimensions}
\label{sec:compare4}

In four dimensions,  SET components depend on the polar angle $\theta $ which complicates the analysis.
We therefore begin by considering the SET in the equatorial plane, $\theta = \pi /2$. 
For this value of $\theta $, the profiles of the energy density $E=3P$ and heat flux $W^{(\varphi )}$ as functions of $\rho $ are very similar to those depicted in Figs.~\ref{fig:3dEnergy} and \ref{fig:3dHeatFlux}  respectively for $n=3$.  
At high temperatures, the differences between the RKT and QFT results for the energy density are even smaller for $n=4$ than for $n=3$. 

\begin{figure*}[t]
\centering
    \includegraphics[width=0.45\linewidth]{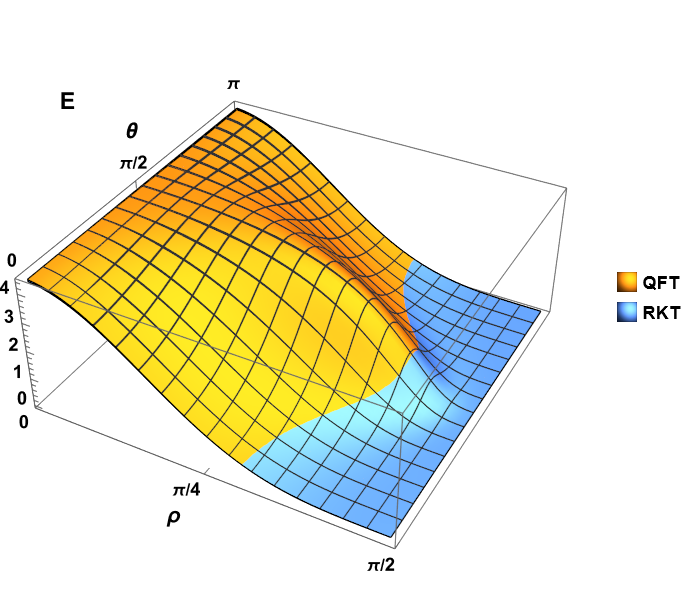}\qquad \qquad
    \includegraphics[width=0.45\linewidth]{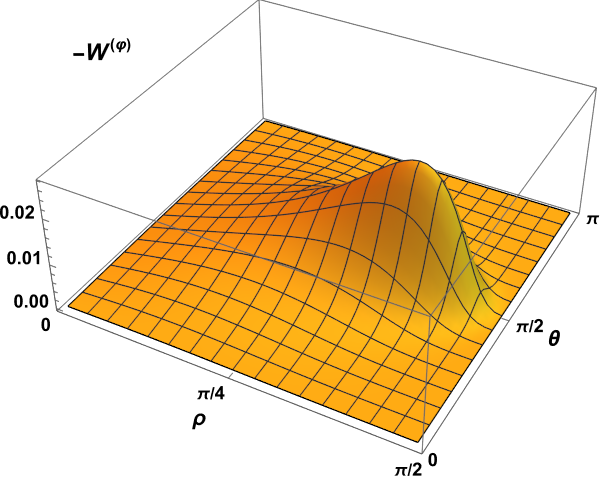}  \\
    \includegraphics[width=0.45\linewidth]{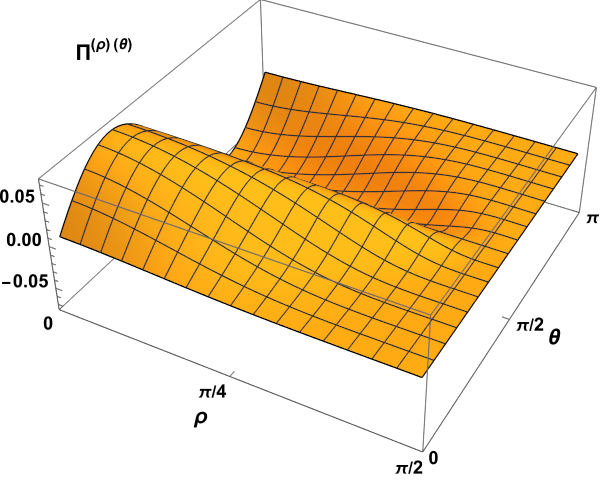} 
    \qquad \qquad
    \includegraphics[width=0.45\linewidth]{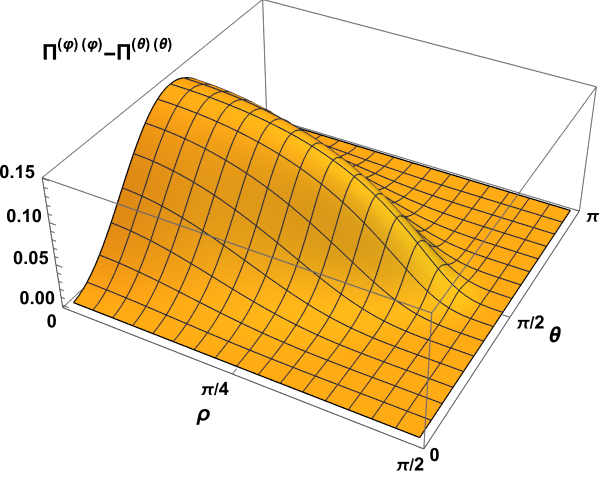}
    \caption{Energy density $E$ (top left), heat flux $-W^{(\varphi )}$ (top right) and the pressure deviator components $\Pi ^{(\rho )(\theta)}$ (bottom left) and  $\Pi ^{(\varphi )(\varphi )}-\Pi ^{(\theta )(\theta )}$ (bottom right) for a rigidly-rotating thermal state on four-dimensional AdS. The inverse temperature is $\beta = \pi /6$ and the angular speed is $\Omega = 0.96$.    
    }
    \label{fig:4dtheta}
\end{figure*}

The profiles of the components of the pressure deviator are rather different for $n=4$ compared to $n=3$.
In Fig.~\ref{fig:4dPresDevRhoPhi} we plot the components $\Pi ^{(\rho )(\rho )}$ (top row) and $\Pi ^{(\varphi )(\varphi )}$ (bottom row) as functions of $\rho $ for $\theta = \pi /2$ and angular speed $\Omega = 0.5$.
Both these quantities vanish on the AdS boundary $\rho \rightarrow \pi/2$, but, unlike the corresponding $n=3$ quantities depicted in Fig.~\ref{fig:3dPresDev}, they do not vanish when $\rho= 0$. 
This is a consequence of the rotation; when $\Omega = 0$, the components of the pressure deviators do vanish at the origin \cite{Ambrus:2018olh,Thompson:2024vuj}. 
In Fig.~\ref{fig:4dPresDevRhoPhi} it can be seen that both $\Pi ^{(\rho )(\rho )}$ and $\Pi ^{(\varphi )(\varphi )}$ have a maximum at the origin. 
While $\Pi ^{(\rho )(\rho )}$ is positive for all values of $\rho $ and decreasing as $\rho $ increases, we find that $\Pi ^{(\varphi  )(\varphi )}$ is negative close to the boundary. 

We further explore how the angular speed $\Omega $ affects the components of the pressure deviator in Fig.~\ref{fig:4dPresDevtheta}, where we have chosen a large value of the temperature $\beta ^{-1}=6/\pi $.
When the angular speed is very small, the components of the pressure deviator vary only a small amount as the polar angle $\theta $ varies, and are negligible at the origin.
When $\Omega = 0$, we have $\Pi ^{(\theta )(\theta ) }=\Pi ^{(\varphi )(\varphi )} = -\Pi ^{(\rho )(\rho )}/2$ \cite{Thompson:2024vuj}, and this remains approximately true for sufficiently small $\Omega $.
As the angular speed $\Omega $ increases, we observe quite different behaviours for these three pressure deviator components. 
At $\rho =0$, the component $\Pi ^{(\rho )(\rho )}$ increases close to the equatorial plane $\theta = \pi /2$ as $\Omega $ increases, but decreases close to the axis of rotation at $\theta = 0,\pi $.
For sufficiently large values of $\Omega $, near the axis, $\Pi ^{(\rho )(\rho )}$ is monotonically increasing as $\rho $ increases, but near the equatorial plane, it is monotonically decreasing as $\rho $ increases. 
The component $\Pi ^{(\theta )(\theta )}$ has the opposite behaviour to $\Pi ^{(\rho )(\rho )}$: at $\rho = 0$ it decreases close to the equatorial plane as $\Omega $ increases, but increases close to the axis of rotation. 
For large $\Omega $, as $\rho $ increases, $\Pi ^{(\theta )(\theta)}$ is monotonically increasing near the equatorial plane but 
monotonically decreasing near the axis of rotation.
In contrast to $\Pi ^{(\rho )(\rho )}$ and $\Pi ^{(\theta )(\theta )}$, a noticeable dependence of $\Pi ^{(\varphi )(\varphi )}$ on the polar angle $\theta $ is apparent only for large values of $\Omega $, in which case $\Pi ^{(\varphi )(\varphi )}$ is monotonically decreasing as $\rho $ increases for all values of $\theta $. 

Finally, in Fig.~\ref{fig:4dtheta}, we consider the dependence of the quantities $E$, $-W^{(\varphi )}$, the remaining nonzero pressure deviator component $\Pi ^{(\rho )(\theta )}$ and the combination $\Pi ^{(\varphi )(\varphi )}-\Pi ^{(\theta )(\theta )}$ on $\rho $ and $\theta $ for $\beta = \pi /6$ and $\Omega = 0.96$.  
These quantities have qualitatively similar properties for other values of the inverse temperature $\beta$
and angular speed $\Omega $.
All except for the energy density $E$ are vanishing when $\Omega = 0$ and in RKT.

In the top-left plot in Fig.~\ref{fig:4dtheta}, we show the energy density $E$ computed in both RKT (blue surface) and QFT (orange surface).  
The RKT energy density is an excellent approximation to that computed in QFT  for the high temperature considered here.  
The QFT energy density is slightly larger than that in RKT for smaller $\rho $, and the RKT energy density is slightly larger than that in QFT close to the AdS boundary.
The dependence of $E$ on the coordinates $(\rho , \theta )$ is qualitatively similar to that of the pressure deviator component $\Pi ^{(\varphi )(\varphi )}$ in the bottom-right plot in Fig.~\ref{fig:4dPresDevtheta} (albeit the energy density is much larger in magnitude). 
These two surfaces mimic the dependence of the local temperature ${\widetilde {\beta }}^{-1}$ (\ref{eq:loctemp}) on $(\rho , \theta )$. 

The heat flux $W^{(\varphi )}$ (top-right plot) vanishes on the axis of rotation and at the AdS boundary. 
As in three dimensions, we find that this quantity is always negative. 
Furthermore, it has a peak on the equatorial plane, the location of the peak moving towards larger $\rho  $ as $\Omega $ increases.
The pressure deviator component $\Pi ^{(\rho )(\theta )}$ (bottom-left plot)  vanishes in the equatorial plane and on the axis of rotation, and is antisymmetric under the mapping $\theta \to \pi - \theta$.
It is notable that, unlike the situation observed here for a quantum scalar field, $\Pi ^{(\rho )(\theta )}$ vanishes identically for a quantum fermion field \cite{Ambrus:2014fka,Ambrus:2021eod}. 
Finally, the combination of pressure deviator components $\Pi ^{(\varphi )(\varphi )}-\Pi ^{(\theta )(\theta )}$ (bottom-right plot) vanishes at the AdS boundary and on the axis of rotation. 
For fixed $\rho $, it has a maximum on the equatorial plane.

\section{Conclusions}
\label{sec:conc}

In this letter we have studied rigidly-rotating thermal states for a massless, conformally-coupled quantum scalar field on three- and four-dimensional AdS. The rate of rotation is assumed to be sufficiently small that there is no speed-of-light surface. 
We have compared the SETs resulting from RKT (modelling the scalar field as a classical gas of bosonic particles) and a full QFT analysis. 
We find that RKT provides an excellent approximation to QFT at high temperatures, for all values of the angular speed. 
At low temperatures, quantum effects become more significant, as previously observed for both scalars \cite{Ambrus:2018olh} and fermions \cite{Ambrus:2017vlf} in nonrotating thermal states on AdS. This result is not unexpected, since a thermal state at high temperature contains a very large number of quantum particles, and is thus well-approximated by a thermal gas of classical particles. 

We find that the profiles of the energy density (and hence the pressure) always have a maximum at the AdS origin, and are monotonically decreasing towards the space-time boundary. 
This is in contrast to the corresponding results for rigidly-rotating states of fermions \cite{Ambrus:2014fka}, where the location of the maximum energy density moves outwards towards the AdS boundary as the angular speed increases.
Therefore rotation has a less dramatic effect on a scalar field compared to a fermion field.
This is due to the nonzero spin of a fermion field coupling to the rigid rotation.

Rigidly-rotating states have more complex SETs than nonrotating states, with a nonzero azimuthal heat flux component, and additional independent pressure deviator components. 
In four dimensions, the local temperature and SET components depend on the polar angle as well as the radial coordinate, introducing further complexity. 
Perhaps surprisingly, the SET for the quantum scalar field considered here has less symmetry than that for the quantum fermion field \cite{Ambrus:2021eod}, with an additional nonzero pressure deviator component. 
We have explored how all these SET components depend on the space-time coordinates, as well as the temperature of the state and the rate of rotation. 
In four dimensions, increasing the angular speed has the greatest effect in the equatorial plane for almost all the quantities studied.

It would be interesting to explore the backreaction of these rigidly-rotating quantum thermal states on the AdS metric.
Using the SET computed in this paper as a source term in the Einstein equations, one could seek rotating quantum-corrected solitons, generalizing the nonrotating quantum-corrected solitons found in \cite{Thompson:2024vuj}.
It would also be of great interest to go beyond the quantum-corrected space-time approximation.
Very recently \cite{Juarez-Aubry:2024slj}, solutions of the full semiclassical Einstein equations on AdS have been found for vacuum states. 
We anticipate that generalizing the results of \cite{Juarez-Aubry:2024slj} to (either rotating or nonrotating) thermal states would be considerably more challenging than the vacuum case, since the resulting space-times will no longer be maximally symmetric. 
We therefore leave such investigations for future work.

\section*{Acknowledgements}
We thank Victor Ambrus for invaluable assistance with the RKT analysis in Sec.~\ref{sec:RKT}, Bruno Felipe for helpful discussions about boundary conditions in QFT, and Francesco Palli for pointing out an error in an earlier version of our {\tt {Mathematica}} notebook.
J.T.~thanks Ethan James German for assistance with {\tt {Mathematica}} computations.
E.W.~thanks Carl Kent for insightful discussions at an early stage in this project.
The work of J.T.~is supported by an EPSRC studentship.
The work of E.W.~is supported by STFC grant number ST/X000621/1.
We acknowledge IT Services at The University of Sheffield for the provision of services for High Performance Computing.
Data supporting this publication can be freely downloaded from the University of Sheffield Research Data Repository at {\url {https://doi.org/10.15131/shef.data.28850714}}, under the terms of the Creative Commons Attribution (CC--BY) licence. 

%\vfill
%\pagebreak

\bibliographystyle{elsarticle-num} 
\bibliography{thermal}

\end{document}